\documentclass[twoside,preprintnumbers,amsmath,amssymb,pacs,shownopacs,nofootinbib]{revtex4}
\usepackage{amsmath,amssymb,url}
\usepackage{graphicx,subfigure}
\usepackage[euler-hat-accent,euler-digits]{eulervm}
\usepackage{braket,euscript,enumitem}
\usepackage{fancyvrb}

\usepackage{amsfonts}

\usepackage{xcolor}
\usepackage[T1,T5]{fontenc}

\newcommand{\omu}{\ensuremath{\bar \mu}}

\newcommand{\MSbar}{\overline{\mbox{MS}}}

\newcommand{\p}{\partial}

\newcommand{\lms}{\Lambda_{\overline{\mbox{\tiny{MS}}}}}
\begin{document}

\title{{\bf The   BRST-invariant   vacuum state of   the   Gribov--Zwanziger theory}}
\author{D.~Dudal$^{\dag,\ddag}$, C.~P.~Felix$^{\dag}$, L.~F.~Palhares$^{\S}$, F.~Rondeau$^{\dag,\star}$, D.~Vercauteren$^+$}
\email{david.dudal@kuleuven.be,caroline.felix@kuleuven.be, leticia.palhares@uerj.br, francois.rondeau@ens-paris-saclay.fr, vercauterendavid@duytan.edu.vn}

\affiliation{$\dag$ KU Leuven Campus Kulak Kortrijk, Department of Physics, Etienne Sabbelaan 53 bus 7657, 8500 Kortrijk, Belgium.\\$^\ddag$ Ghent University, Department of Physics and Astronomy, Krijgslaan 281-S9, 9000 Gent, Belgium\\ $^\S$ Instituto de F\'isica Te\'orica, Rua S\~ao Francisco Xavier 524, 20550-013, Maracan\~a, Rio de Janeiro, Brasil\\$\star$Ecole Normale Sup\'erieure Paris-Saclay, Avenue du Pr\'esident Wilson 61, 94235 Cachan Cedex, France \\ $^+$ Duy T\^an University, Institute of Research and Development, P809, 3 Quang Trung, {\fontencoding{T5}\selectfont H\h ai Ch\^au, \DJ \`a N\~\abreve ng}, Vietnam}


\begin{abstract}
We revisit the effective action of the Gribov--Zwanziger theory, taking into due account the BRST symmetry and renormalization (group invariance) of the construction. We compute at   one loop   the effective potential, showing the emergence of   BRST-invariant   dimension 2 condensates stabilizing the vacuum. This paper sets the stage at zero temperature, and clears the way to studying the Gribov--Zwanziger gap equations, and particularly the horizon condition, at finite temperature in future work.
\end{abstract}
\maketitle

\section{Introduction}

Up until now, quark and gluon confinement has not been   rigorously   proven. It is well known that the perturbative formalism fails for non-Abelian gauge theories at low energy, since the coupling constant $g^2$ is strong. To get reliable results in the infrared (IR) in the continuum formulation, non-perturbative methods are needed. For an overview of such methods and obtained results, let us refer for example to
\cite{Roberts:1994dr,Alkofer:2000wg,Aguilar:2004sw,Dudal:2005na,Fischer:2006ub,Gracey:2006dr,Dudal:2007cw,Dudal:2008sp,Aguilar:2008xm,Fischer:2008uz,Boucaud:2008ky,Binosi:2009qm,Tissier:2010ts,Gracey:2010cg,Tissier:2011ey,Bashir:2011dp,Maas:2011se,Dudal:2011gd,Boucaud:2011eh,Boucaud:2011ug,Vandersickel:2012tz,Aguilar:2012rz,Cucchieri:2012cb,Ayala:2012pb,Bashir:2012fs,Huber:2012kd,Serreau:2012cg,Rojas:2013tza,Pelaez:2013cpa,Eichmann:2014xya,Aguilar:2014lha,Reinosa:2014zta,Cyrol:2014kca,Blum:2014gna,Huber:2015ria,Capri:2015ixa,Capri:2015nzw,Siringo:2015gia,Aguilar:2015nqa,Binosi:2016wcx,Aguilar:2016vin,Eichmann:2016yit,Capri:2016aqq,Capri:2016aif,Capri:2016gut,Cyrol:2016tym,Pereira:2016inn,Reinosa:2017qtf,Capri:2017abz,Mintz:2017qri,Bermudez:2017bpx,Binosi:2017rwj,Cyrol:2017ewj,Comitini:2017zfp,Mintz:2018hhx,Huber:2018ned,Siringo:2018uho,Capri:2018ijg}. Notice that the continuum formulation requires gauge fixing, in which case lattice analogues of dedicated gauge fixings can be a powerful ally giving complementary insights, see \cite{Boucaud:2001st,Bowman:2007du,Cucchieri:2007md,Sternbeck:2007ug,Bogolubsky:2007ud,Cucchieri:2008fc,Boucaud:2008gn,Cucchieri:2008qm,Maas:2008ri,Bornyakov:2009ug,Cucchieri:2009kk,Cucchieri:2011ig,Oliveira:2012eh,Bornyakov:2013pha,Bornyakov:2013ysa,Maas:2015nva,Bicudo:2015rma,Duarte:2016iko,Dudal:2018cli,Cucchieri:2018doy,Boucaud:2018xup} for some relevant works in this area.

Motivated by this, a number of studies over the past decade have focused on the gluon, quark and also ghost propagator in the infrared region, where color degrees of freedom are confined. Although these objects are unphysical by themselves --- being gauge variant --- they are nevertheless the basic building blocks, next to the interaction vertices, entering   gauge-invariant   objects directly linked to physically relevant quantities such as the spectrum, decay constants, critical exponents and temperatures, etc.

One particular way to deal with non-perturbative physics at the level of elementary degrees of freedom is by dealing with the Gribov issue \cite{Gribov:1977wm,Vandersickel:2012tz}: the fact that there is no unique way of selecting one representative configuration of a given gauge orbit in covariant gauges \cite{Singer:1978dk}. As there is also no rigourous way to deal properly with the existence of gauge copy modes in the path integral quantization procedure, in this paper we will use  a well-tested  formalism available to deal with the issue, which is known as the Gribov--Zwanziger (GZ) formalism: a restriction of the path integral to a smaller subdomain of gauge fields \cite{Gribov:1977wm,Zwanziger:1989mf,Zwanziger:1992qr}.

This approach was first proposed for the Landau and the Coulomb   gauges  . It long suffered from a serious drawback: its concrete implementation seemed to be inconsistent with BRST (Becchi--Rouet--Stora--Tyutin \cite{Becchi:1974xu,Becchi:1975nq,Tyutin:1975qk}) invariance of the gauge-fixed theory, which clouded its interpretation as a gauge (fixed) theory. Only more recently was it realized by some of us and colleagues how to overcome this complication to get a   BRST-invariant restriction of the gauge path integral  . As a bonus, the method also allowed the generalization of the GZ approach to the linear covariant gauges, amongst others \cite{Capri:2015ixa,Capri:2015nzw,Capri:2016aqq,Capri:2016gut}.

Another issue with the original GZ approach was that some of its major   leading-order   predictions did not match the corresponding lattice output. Indeed, in   the    case of the Landau gauge, the GZ formalism by itself predicts   at tree level   a gluon propagator vanishing at momentum $p=0$, next to, more importantly, a ghost propagator with a stronger than $1/p^2$ singularity for $p\to0$. Although the latter fitted well in the Kugo--Ojima confinement criterion \cite{Kugo:1979gm}, it was at odds with large volume lattice simulations \cite{Cucchieri:2007md,Sternbeck:2007ug}. By now, several analytical takes exist on this, all being compatible, qualitatively and/or quantitatively, with lattice data, not only for elementary propagators but also for vertices \cite{Dudal:2007cw,Dudal:2008sp,Aguilar:2008xm,Fischer:2008uz,Boucaud:2008ky,Binosi:2009qm,Tissier:2010ts,Gracey:2010cg,Tissier:2011ey,Bashir:2011dp,Maas:2011se,Dudal:2011gd,Boucaud:2011eh,Boucaud:2011ug,Vandersickel:2012tz,Aguilar:2012rz,Cucchieri:2012cb,Ayala:2012pb,Bashir:2012fs,Huber:2012kd,Serreau:2012cg,Rojas:2013tza,Pelaez:2013cpa,Eichmann:2014xya,Aguilar:2014lha,Reinosa:2014zta,Cyrol:2014kca,Blum:2014gna,Huber:2015ria,Capri:2015ixa,Capri:2015nzw,Siringo:2015gia,Aguilar:2015nqa,Binosi:2016wcx,Aguilar:2016vin,Eichmann:2016yit,Capri:2016aqq,Capri:2016aif,Capri:2016gut,Cyrol:2016tym,Pereira:2016inn,Reinosa:2017qtf,Capri:2017abz,Mintz:2017qri,Bermudez:2017bpx,Binosi:2017rwj,Cyrol:2017ewj,Comitini:2017zfp,Mintz:2018hhx,Huber:2018ned,Siringo:2018uho}.  In the GZ formalism, in particular,   the situation can be remedied by correctly incorporating the effects of certain mass dimension two condensates, the importance of which was already stressed before in papers like \cite{Gubarev:2000eu,Gubarev:2000nz,Boucaud:2001st,Verschelde:2001ia,Kondo:2001nq}. This idea was first put on the table in \cite{Dudal:2007cw,Dudal:2008sp} and later on a self-consistent computational scheme was constructed in \cite{Dudal:2011gd} based on the effective action formalism for local composite operators developed in \cite{Verschelde:1995jj,Verschelde:2001ia}, the renormalization of which was proven in \cite{Dudal:2002pq}. Unfortunately,   the explicit computation of the effective action was not achieved at the time, while the setup was still based on the   BRST-breaking   GZ proposal.

The goal of this paper is thus to revisit, in the newly established   BRST-invariant   setting, the dynamical generation of $d=2$ condensates, the latter themselves affiliated to   BRST-invariant   operators. Said otherwise, we will explicitly construct the non-perturbative GZ vacuum, which will be shown to have a lower vacuum energy compared to the original GZ action.   Moreover, we show that the original action represents a totally unstable point of the effective potential, while the formation of the condensates properly produces a minimum.   The GZ vacuum thus stabilizes itself by the formation of non-trivial condensates, which in return affect the dynamics of the field excitations above that vacuum. The practical problems to compute the effective potential that plagued \cite{Dudal:2011gd} are circumvented here by a clever use of Hubbard--Stratonovich transformations.

This paper is organized as   follows.   In Section \ref{GZ action}, we briefly introduce the    BRST-invariant   Gribov--Zwanziger formalism for the class of linear covariant gauges. The transition from the Gribov--Zwanziger to the Refined Gribov--Zwanziger (RGZ) procedure is described in Section \ref{RGZ formalism}. We also concisely explain the renormalization group equation aspects of the effective action construction for a set of $d=2$ BRST invariant local composite operators in Section \ref{RGE section}. In the Section \ref{work}, the one-loop calculation of the effective potential is presented. Finally, in Section \ref{solve}, the physical solution is identified in $\MSbar$ and in general schemes.

\section{The   BRST-invariant   Gribov--Zwanziger action in linear covariant gauges}\label{GZ action}
It is well known that at low energy, we have to deal with Gribov copies, in principle both with ``large'' and infinitesimal ones. In the low-energy regime, such copies are not suppressed because the coupling constant $g$ is large \cite{Gribov:1977wm}. A way to avoid the ambiguity --- or at least the ambiguity coming from the infinitesimal ones --- is to restrict the functional integral over the gauge fields to a specific region $\Omega$ in field space where no infinitesimal Gribov copies exist --- as was originally proposed by Gribov in the Landau gauge \cite{Gribov:1977wm}. As the Gribov ambiguity exists for any covariant gauge \cite{Singer:1978dk}, it will in particular be present in the class of widely used linear covariant gauges, to which Feynman gauge and Landau gauge belong. It was only recently discussed how to treat these copies in linear covariant gauges other than the Landau gauge \cite{Capri:2015ixa,Capri:2015nzw,Capri:2016aqq,Capri:2016aif,Capri:2016gut}.

The construction eliminating (infinitesimal) Gribov copies in general linear covariant gauges is based on the field $A_\mu^h$, which is the gauge transformed configuration of $A_\mu$ minimizing the functional
\begin{equation} \begin{gathered}
	f_A[u] \equiv \min_{\{u\}} \operatorname{Tr} \int d^{ d}x\,A_{\mu}^{u}A_{\mu }^{u} \;, \\
	A_{\mu }^{u} =u^{\dagger }A_{\mu }u+\frac{i}{g}u^{\dagger }\partial _{\mu}u \;,  \label{Aminn0}
\end{gathered} \end{equation}
which is obtained through iterative minimization of the functional $f_A[u]$ along the gauge orbit of $A_{\mu}$ \cite{Dell'Antonio:1989jn,vanBaal:1991zw,Lavelle:1995ty}. The field $A_\mu^h$ is a non-local power series in the gauge field; iterative minimization produces the following local minimum:
\begin{subequations} \begin{gather}
	A_{\mu }^{h} =\left( \delta _{\mu \nu }-\frac{\partial _{\mu }\partial_{\nu }}{\partial ^{2}}\right) \phi _{\nu }\;,  \qquad  \partial_\mu A^h_\mu= 0 \;, \\
	\phi _{\nu } =A_{\nu }-ig\left[ \frac{1}{\partial ^{2}}\partial A,A_{\nu}\right] +\frac{ig}{2}\left[ \frac{1}{\partial ^{2}}\partial A,\partial_{\nu }\frac{1}{\partial ^{2}}\partial A\right] + \mathcal O(A^{3}) \;.
\end{gather} \label{min0} \end{subequations}
It is worth pointing out that the quantity $A_\mu^h$ is gauge invariant order by order \cite{Capri:2015ixa,Capri:2015nzw,Capri:2016aqq,Capri:2016aif,Capri:2016gut}. If we couple $A_\mu^h$ to the Yang--Mills action in a general linear covariant gauge, it seems this will result in a non-local quantum field theory. Fortunately, the field $A_\mu^h$ can be localized by adding an auxiliary Stueckelberg field $\xi^a$ \cite{Capri:2015ixa,Capri:2015nzw,Capri:2016aqq,Capri:2016aif,Capri:2016gut,Fiorentini:2016rwx} so that
\begin{equation}
	A^{h}_{\mu}=(A^{h})^{a}_{\mu}T^{a}=h^{\dagger}A^{a}_{\mu}T^{a}h+\frac{i}{g}\,h^{\dagger}\partial_{\mu}h \;, \qquad h=\textrm{e}^{ig\,\xi^{a}T^{a}} \;,
\end{equation}
and by imposing that $A^h_\mu$ is transverse, $\p_\mu A_\mu^h=0$. Now, the local gauge invariance of $A^{h}_{\mu}$ under a gauge transformation $u\in SU(N)$ can be appreciated from
\begin{equation}
h\to u^\dagger h\;,\qquad\ h^\dagger\to h^\dagger u\;,\qquad A_\mu \to u^\dagger A_\mu u + \frac{i}{g}u^\dagger \p_\mu u \;.
\end{equation}

Using this field $A_\mu^h$, a Gribov region $\Omega$ not containing any infinitesimal Gribov copies is given by
\begin{equation}
	\Omega=\{A_{\mu}^a;\  \partial_\mu A_{\mu}^a=i\alpha b^a,\qquad \mathcal{M}^{ab}(A^h)=-\partial_\mu D_\mu^{ab}(A^h)>0 \} \;,
	\label{GR}
\end{equation}
where a Hermitian Faddeev--Popov-like operator\footnote{This is not the Faddeev--Popov operator for a generic linear covariant gauge, the latter is given by the non-Hermitian operator $-\p D(A)$.}, $\mathcal{M}^{ab}(A^h)=-\delta^{ab}\partial^2+gf^{abc}(A^h)_{\mu}^{c}\partial_{\mu}$, is required to be positive. Implementing the positivity of the Hermitian operator $-\p D (A^h)$ is a sufficient condition to kill off a large set of gauge copies in linear covariant gauges, namely those that are continuously connected to infinitesimal copies in Landau gauge, as has been discussed in \cite{Capri:2015ixa}. More precisely, we impose that the Fourier transform of the inverse operator of $-\p D (A^h)$ displays no poles for $p^2>0$. This constraint can, in the thermodynamic limit, be lifted into the path integral using a saddle point evaluation. The saddle point equation is nothing else than the horizon condition, which in its original, non-local, form reads   in $d$ dimensions
\begin{equation}\label{gapnonl}
	\Braket{h(x)}=d(N^2-1)\,,\qquad h(x) = g^2\gamma^4\int d^d x f^{akc} A_\mu^{h,k}(x) \left[-\partial_\mu D_\mu^{ab}(A^h)\right]^{-1}_{(x,y)} f^{b m c} A_\mu^{h,m}(y)
\end{equation}
We refer to \cite{Capri:2015ixa,Capri:2015nzw} for the detailed derivation, see also \cite{Gribov:1977wm,Zwanziger:1989mf,Zwanziger:1992qr,Vandersickel:2012tz,Capri:2012wx}.

The total action implementing the Gribov horizon condition in a general linear covariant gauge is given by
\begin{subequations} \begin{equation}\label{action}
	S= S_{YM} + S_{GF} + S_{GZ} + S_{\varepsilon} \;.
\end{equation}
In this expression, $S_{YM}$ is the Yang--Mills action
 \begin{equation}
S_{YM}  = \frac{1}{4} \int d^{ d}x  F^a_{\mu\nu} F^a_{\mu\nu} \;; \label{ym}
\end{equation}
	$S_{GF}$ denotes the Faddeev--Popov gauge fixing in the linear covariant gauge:
\begin{equation}
	S_{GF} = \int d^{ d}x  \left( \frac{\alpha_g}{2}\,b^{a}b^{a}+ib^{a}\,\partial_{\mu}A^{a}_{\mu}+\bar{c}^{a}\partial_{\mu}D^{ab}_{\mu}(A)c^{b}     \right) \;, \label{sfpb}
\end{equation}
with $\alpha_g$ the gauge parameter, which is zero for the Landau gauge; and $S_{GZ}$ is the Gribov--Zwanziger action in its local form, which can be written as
\begin{eqnarray}
	S_{GZ}&=&\int d^{ d}x\left[\bar{\varphi}^{ac}_{\mu}\partial_{\nu}D^{ab}_{\nu}(A^h)\varphi^{bc}_{\mu}-\bar{\omega}^{ac}_{\mu}\partial_{\nu}(D^{ab}_{\nu}(A^h)\omega^{bc}_{\mu})+\bar{\eta}^{a}\partial_{\mu}D^{ab}_{\mu}(A^h)\eta^{b}\right]
\nonumber\\&&-\gamma^{2}g\int d^4 x\left[f^{abc}(A^h)^{a}_{\mu} (\varphi^{bc}_{\mu}+\bar{\varphi}_{\mu}^{bc}) + \frac{d}{g}( N^{2}  -1)\gamma^{2}\vphantom{\frac{1}{2}}\right] \;, \label{S_gamma2}
\end{eqnarray} \end{subequations}
The localizing fields $(\bar{\varphi}^{ac}_{\mu},~\varphi^{ac}_{\mu})$ are a pair of complex-conjugate bosonic fields, while $(\bar{\omega}^{ac}_{\mu},~\omega^{ac}_{\mu})$ a pair of anti-commuting complex-conjugate fields. The fields $\bar\eta^a$ and $\eta^a$ are also ghost-like, while $\gamma$ is the Gribov parameter, which is dynamically fixed by a gap equation \cite{Zwanziger:1989mf,Zwanziger:1992qr,Capri:2012wx,Capri:2015nzw},
\begin{equation}\label{gap}
	\braket{f^{abc}(A^h)_\mu^a({\varphi}_{\mu}^{bc}+\bar{\varphi}_{\mu}^{bc})}=2d(N^2-1)\frac{\gamma^2}{g^2},
\end{equation}
also known as the horizon condition. This equation can be succinctly rewritten as
\begin{equation}\label{gapG}
	\frac{\p \Gamma}{\p \gamma^2}=0 \;,
\end{equation}
where $\Gamma$ is the quantum action defined by
\begin{equation}\label{quantumAction}
	\textrm{e}^{-\Gamma}=\int[d\Phi ] \textrm{e}^{-S}
\end{equation}
with $[d\Phi]$ the Haar measure of integration over all the quantum fields present in the action.

Finally, the  term $S_{\varepsilon}$
\begin{equation}
S_{\varepsilon} = \int d^{  d}x\; \varepsilon^{a}\,\partial_{\mu}(A^h)^{a}_{\mu}   \label{stau}
\end{equation}
implements, through the Lagrange multiplier $\varepsilon$, the transversality of the composite operator $(A^h)_\mu^a$,  namely $\partial_{\mu}(A^h)^{a}_{\mu}=0$.

The action $S$ in eq.~\eqref{action} enjoys an exact BRST invariance, $s S = 0$ and $s^2=0$,  expressed by \cite{Capri:2015ixa,Capri:2015nzw,Capri:2016aqq,Capri:2016aif,Capri:2016gut}
\begin{equation} \begin{gathered}
s A^{a}_{\mu}=-D^{ab}_{\mu}c^{b} \;,\\
s c^{a}=\frac{g}{2}f^{abc}c^{b}c^{c}\;, \qquad s \bar{c}^{a}=ib^{a}\;, \\
s b^{a}= 0\;, \\
s \varphi^{ab}_{\mu}= 0 \;, \qquad s \omega^{ab}_{\mu}=0\;, \\
s\bar\omega^{ab}_{\mu}=0  \;, \qquad s\bar\varphi^{ab}_{\mu}=0\;, \\
s\varepsilon^{a}=0    \;, \qquad s (A^h)^{a}_{\mu}= 0 \;, \\
s h^{ij} = -ig c^a (T^a)^{ik} h^{kj} \;.
\end{gathered} \label{brstgamma} \end{equation}
Notice that the gap equation \eqref{gap} is a   BRST-invariant    condition. The multiplicative renormalizability of this construction was proven, to all orders, in \cite{Capri:2017npq,Capri:2017bfd}.

\section{Refined Gribov--Zwanziger Action}\label{RGZ formalism}
In \cite{Dudal:2007cw}, it was noticed that the GZ formalism  in Landau gauge is plagued by non-perturbative dynamical instabilities, leading to the formation of $d=2$ condensates like $\langle A_{\mu}^{a}A_{\mu}^{a}\rangle$ and $\langle \bar{\varphi}^{ab}_{\mu}\varphi^{ab}_{\mu}-\bar{\omega}^{ab}_{\mu}\omega^{ab}_{\mu} \rangle$, which are energetically favored \cite{Dudal:2007cw,Dudal:2008sp,Dudal:2011gd}. Later, similar features were noticed in the Maximal Abelian gauge GZ formulation \cite{Capri:2015pfa,Capri:2017abz}.  This led to the Refined Gribov--Zwanziger formalism, which explicitly takes into account the effects of these condensates.

In this paper, we will work out in detail the dynamical RGZ formalism in linear covariant gauges. In order to do so, we will couple the   BRST-invariant   operators $A_\mu^{h,a} A_\mu^{h,a}$ and $\bar{\varphi}^{ab}_{\mu}\varphi^{ab}_{\mu}$ to the GZ action via the local composite operator (LCO) formalism. As a final result, the Refined Gribov--Zwanziger (RGZ) action \eqref{Final W} will be obtained. With this RGZ action, the dominant IR ghost behavior is $1/p^2$, while the gluon propagator, at tree-level but in the new improved vacuum, is given by
\begin{equation}
\Braket{A^a_\mu(p)A^b_\nu(-p)}=\frac{p^2+M^2}{p^4+(M^2+m^2)p^2+M^2m^2+\lambda^4}     {\mathcal P}_{\mu\nu}(p)   \delta^{ab}+\frac{\alpha_g}{p^2}L_{\mu\nu}\delta^{ab} \;,
\label{gluon_prop_faynman}
\end{equation}
where
\begin{equation}
  {\mathcal P}_{\mu\nu}(p)   =\delta_{\mu\nu}-\frac{p_{\mu}p_{\nu}}{p^{2}} \;, \qquad L_{\mu\nu}=\frac{p_\mu p_\nu}{p^2} \;,
\label{tranverse_prop}
\end{equation}
are the transversal and longitudinal projectors, $\lambda^4=2g^2N\gamma^4$, and $M^2$ and $m^2$ are the mass scales linked to the condensates $\langle \bar{\varphi}^{ab}_{\mu}\varphi^{ab}_{\mu}\rangle$ and $\langle A_{\mu}^{h,a}A_{\mu}^{h,a}\rangle$, respectively  (see later). It can be shown, \cite{Capri:2016aqq}, that the longitudinal form factor remains bare, as is usual in the linear covariant gauge. This fact is also confirmed non-perturbatively using lattice simulations \cite{Cucchieri:2009kk,Bicudo:2015rma} and is consistent with the findings in \cite{Huber:2015ria,Aguilar:2015nqa} as well.

For later usage, we remind here that, using the Nielsen identities, it can be shown that the poles of the gluon propagator are gauge parameter and renormalization scale independent order per order, even in the GZ case. See the detailed discussion in \cite{Capri:2016gut}. Evidently, BRST invariance is crucial here as this underlies the Nielsen identities. We will later on use this knowledge.

Depending on the relative size of the mass scales appearing in \eqref{gluon_prop_faynman}, the propagator can  develop complex-conjugate  poles. If \eqref{gluon_prop_faynman} is fitted to lattice data, the complex pole scenario is clearly preferred \cite{Dudal:2010tf,Cucchieri:2011ig,Dudal:2018cli}. These complex poles evidently remove the gluon from the physical spectrum, which  could offer  an intuitive explanation of why gluons are unobservable. Notice that these complex poles also occur explicitly in other approaches, see \cite{Comitini:2017zfp,Hayashi:2018giz}.

To compute the effective potential of the above-mentioned condensates, we add the local sources $\tau$ and $Q$, coupled to the relevant local composite operators, to the action $S$ given in \eqref{action}:
\begin{subequations} \begin{equation}
  \Sigma=S+S_{A^2}+S_{\varphi\bar{\varphi}}+S_{\rm{vac}} \;.
  \label{action_condensate}
\end{equation}
In this expression, we have, including the $Z$-factors in the conventions of \cite{Dudal:2011gd}, that
\begin{align}
S_{A^2}&=\int d^d x Z_A (Z_{\tau\tau}\tau + Z_{\tau Q}Q) \frac12A^{h,a}_\mu A^{h,a}_\mu \;, \\
S_{\bar{\varphi}\varphi}&=\int d^d x Z_{QQ} Z_\varphi Q\bar{\varphi}^{ac}_\mu \varphi^{ac}_\mu \;, \\
S_{\rm{vac}}&= -\int d^d x \left(\frac{Z_\zeta\zeta}{2}\tau^2 + Z_\alpha\alpha Q^2 + Z_\chi\chi Q\tau\right) \label{svac} \;.
\end{align} \label{actions} \end{subequations}
In the above expressions, we already used the fact that the source $Q$ has no mixing with $\tau$ (i.e.~$Z_{Q\tau}=0$), while $\tau$ does mix with $Q$, see later. At the operator level, this means $\bar\varphi \varphi$ mixes with $A^h A^h$, while $A^h A^h$ renormalizes on its own. The sources are BRST singlets,
\begin{equation}
s\tau=0, \qquad sQ=0.
\end{equation}

When computing the generating functional, new divergences proportional to $\tau^2$, $Q^2$ and $\tau Q$ appear. This happens because of the divergences appearing in correlation functions such as $\Braket{\mathcal{O}_j(x)\mathcal{O}_j(y)}$, with $\mathcal{O}_i$ one of the $d=2$ operators added to the RGZ action. This is why the term $S_{\rm{vac}}$ given in \eqref{svac} is necessary. The counterterms, which come with new and \emph{a priori} free parameters $\alpha$, $\chi$ and $\zeta$ (so-called LCO parameters), will absorb the divergences in $\tau^2$, $Q^2$ and $Q\tau$, \emph{i.e.}~via $\delta \zeta \tau^2$, $\delta\alpha Q^2$ and $\delta\chi Q\tau$. We will momentarily discuss how to fix the (finite) parameters themselves, while maintaining full multiplicative renormalizability. This method was originally developed in \cite{Verschelde:1995jj}, see also \cite{Verschelde:2001ia,Knecht:2001cc}. The generalization, including operator mixing, was worked out first in \cite{Dudal:2011gd}, and we will rely on the latter reference.

Given that the main purpose of this work is to compute $d=2$ vacuum condensates which are BRST invariant, we can actually make use of the full power of BRST. Indeed, we can choose an appropriate gauge for explicit computation. Clearly, the Landau gauge is singled out, as in that case $A^h A^h$ collapses into $A^2$. Loosely speaking, this is clear from expression \eqref{min0}. A more formal proof based on integrating over the auxiliary fields $\xi$, $\epsilon$, $\eta$ and $\bar\eta$ is provided in \cite{Capri:2016gut}, establishing that the BRST invariant action for $\alpha_g\to0$ exactly reduces to that of the original GZ action in Landau gauge (modulo the extra $d=2$ operators of course).

As such, we can rely on the algebraic renormalization analysis already performed in \cite{Dudal:2011gd}, establishing the renormalizability of the action to all orders of perturbation theory. Moreover, it was shown that the sources $(Q,\tau)$ have the following renormalization structure
\begin{equation}
  \begin{pmatrix}
     Q_0 \\
     \tau_0
  \end{pmatrix} =
          \begin{pmatrix}
            Z_{QQ} &0 \\
           Z_{\tau Q} &  Z_{\tau\tau}          \end{pmatrix}
\begin{pmatrix}
    Q\\
    \tau
  \end{pmatrix} \;.
\label{matrix} \end{equation}

\section{Essential points of the LCO formalism}\label{RGE section}
This section is largely based on \cite[Sect.~4.1]{Dudal:2011gd}. In order to make the paper self-contained, we now review the main steps.

We are interested in the generating functional
\begin{equation}\label{W}
e^{-\Gamma(J)}=\int [d\Phi] e^{-\Sigma}
\end{equation}
where $J= \begin{pmatrix} Q \\ \tau \end{pmatrix}$ and the classical action, with sources, has been written down in \eqref{action_condensate}. At the bare level and in dimensional regularization ($d=4-\epsilon$), we have
\begin{equation}\label{notatie}
-\frac{1}{2} \zeta_0 \tau_0^2 - \alpha_0 Q_0^2 - \chi_0 Q_0 \tau_0  = -\mu ^{-\epsilon} \left( \frac{1}{2} \zeta \tau^2 + \alpha Q^2 + \chi Q \tau + \frac{1}{2} \delta \zeta \tau^2 + \delta \alpha Q^2 +\delta \chi Q \tau \right)\;,
\end{equation}
where we used $\delta\zeta$, $\delta\alpha$ and $\delta\chi$ to denote the corresponding vacuum counterterms \cite{Dudal:2011gd}, necessary to remove the divergences in the sources squared that arise when computing the generating functional. We also already introduced the renormalization scale $\mu$ necessary for dimensional reasons.

The renormalization matrix can be translated into an anomalous dimension matrix $\gamma$ \cite{Dudal:2011gd},
\begin{equation}\label{gammas}
\gamma =  \begin{pmatrix}
              Z_{QQ}^{-1} \mu \frac{\p}{\p \mu} Z_{QQ} & 0 \\
              -Z_{\tau Q} \mu \frac{\p}{\p \mu} Z_{QQ} + Z_{\tau \tau}^{-1} \mu \frac{\p}{\p \mu} Z_{\tau Q} &  Z_{\tau\tau}^{-1} \mu \frac{\p}{\p \mu} Z_{\tau \tau}          \end{pmatrix} =  \begin{pmatrix}
              \gamma_{QQ} & 0 \\
              \gamma_{21} &  \gamma_{\tau \tau}          \end{pmatrix}\;.
\end{equation}
so that
\begin{equation}
\mu \frac{\p}{\p \mu} J = -\gamma\cdot J \;.
\end{equation}
Next, deriving \eqref{notatie} w.r.t.~$\mu$ and identifying terms in $Q^2$,$\tau^2$ and $Q\tau$, we find 3 coupled differential equations
\begin{align}\label{diff1}
&  \beta(g^2)  \frac{\p}{\p g^2} \frac{\zeta (g^2)}{2}  =  \frac{\epsilon}{2}  \delta \zeta  - \frac{1}{2}\beta(g^2)  \frac{\p}{\p g^2} (\delta \zeta)  +  \gamma_{\tau \tau}(g^2) (\zeta + \delta \zeta) \;,\nonumber\\
& \beta(g^2)  \frac{\p}{\p g^2} \alpha (g^2)  =  \epsilon \delta \alpha  - \beta(g^2)  \frac{\p}{\p g^2} (\delta \alpha)  + 2  \gamma_{QQ}(g^2) (\alpha + \delta \alpha) + \Gamma_{21}(g^2) (\chi + \delta \chi) \;, \nonumber\\
& \beta(g^2)  \frac{\p}{\p g^2} \chi (g^2) =  \epsilon  \delta \chi  - \beta(g^2)  \frac{\p}{\p g^2} (\delta \chi)  +   \gamma_{QQ}(g^2) (\chi+ \delta \chi) + \gamma_{\tau \tau}(g^2) (\chi + \delta \chi)  + \Gamma_{21} (g^2) (\zeta + \delta \zeta) \;.
\end{align}
where, following the LCO formalism \cite{Verschelde:1995jj}, we made $\zeta(g^2)$, $\alpha(g^2)$ and $\xi(g^2)$ functions of $g^2$, such that they are no longer free parameters but completely determinable by solving the previous renormalization-group based equations. In practice, this happens order per order in perturbation theory by using a Laurent expansion in $g^2$. This choice (which is unique, see \cite{Verschelde:1995jj,Knecht:2001cc}) is compatible with multiplicative renormalizability of the parameters, in addition to ensuring a homogeneous renormalization group equation of the standard type for the generating functional,
\begin{equation}\label{rggg}
  \left(\mu\frac{\p}{\p\mu}+\beta(g^2)\frac{\p}{\p g^2}+ \int d^dx J\cdot\gamma\cdot\frac{\delta}{\delta J}\right)\Gamma(J)=0.
\end{equation}
Note that in deriving the relations \eqref{diff1}, the finiteness of $(\zeta,\alpha,\xi)$ plays a role.

\section{Computation of The Effective Action }\label{work}
Notice that the action $\Sigma$ in \eqref{action_condensate}  has
three terms quadratic in the sources. These terms introduce a conceptual difficulty:
the interpretation of the effective action $\Gamma$ as an energy density. Indeed, when the sources $J$ are linearly coupled to fields $\sigma$, the functional $\Gamma(J)$ can be Legendre transformed into $\Gamma(\sigma)$. However, if $\Gamma(J)$ contains squares (or higher powers) of the sources, these terms would not cancel out in the Legendre transform, such that the interpretation of $\Gamma$ as an energy density is unclear.

In \cite{Verschelde:2001ia,Verschelde:1995jj}, it was shown how to circumvent this apparent problem by a suitable Hubbard--Stratonovich transformation. In the case of mixing sources/operators, a generalization of this strategy was first worked out in \cite{Dudal:2011gd}. Here, we will use a slightly different version from that of \cite{Dudal:2011gd}, which offers the advantage that --- despite the observations in \cite{Verschelde:2001ia,Verschelde:1995jj} --- it is not necessary to perform $(n+1)$-loop computations to get $n$-loop results with the LCO formalism. That this is possible was first noticed in \cite{Lemes:2002rc}.

To get rid of these quadratic terms in the sources, we proceed by introducing two auxiliary fields $\sigma_1$ and $\sigma_2$ through two identities
\begin{subequations} \label{HS transfo} \begin{gather}
1 = \int [\mathcal{D} \sigma_1] \ e^{- \frac{1}{2Z_\zeta} \int d^d x \left(\sigma_1 + \frac{\bar{a}}{2} A^2 + \bar{b} Q + \bar{c}\tau \right)^2}, \\
1 = \int [\mathcal{D} \sigma_2] \ e^{+ \frac{1}{2Z_\alpha} \int d^d x \left(\sigma_2 + \bar{d} \overline{\varphi} \varphi + \bar{e} Q + \frac{\bar{f}}{2} A^2 \right)^2}, \label{poshs}
\end{gather} \end{subequations}
with which we multiply the integral in \eqref{W}. The positive sign in the exponent of the second integral \eqref{poshs} obviously makes it infinite. However, we should remind that all functional integrals are actually defined only up to an infinite constant, often not explicitly written. Actually, what we are doing by inserting this ``infinite identity'' can be seen as a rescaling of the infinite constant hidden in expression \eqref{W}. Doing this rescaling in a subtle way, a careful choice of the coefficients $\bar a$ to $\bar f$ allows us to eliminate all the quadratic terms in sources appearing in the partition function.

A straightforward computation shows that we have to choose the coefficients
\begin{subequations} \begin{eqnarray}
\bar{a} &=& \frac{Z_AZ_{\tau\tau}}{\sqrt\zeta} \mu^{\epsilon/2} \\
\bar{b} &=& -\frac{Z_\chi \chi}{\sqrt\zeta} \mu^{- \epsilon/2}, \\
\bar{c} &=&  -Z_\zeta \sqrt\zeta \mu^{- \epsilon/2}, \\
\bar{d} &=& \frac{Z_{\varphi}Z_{QQ}}{\sqrt{-2\alpha + \frac{Z_\chi^2\chi^2}{Z_\alpha Z_\zeta\zeta}}}\mu^{\epsilon/2},\\
\bar{e} &=& Z_\alpha \sqrt{-2\alpha + \frac{Z_\chi^2\chi^2}{Z_\alpha Z_\zeta\zeta}} \ \mu^{-\epsilon/2}, \\
\bar{f} &=& -\frac{\frac{Z_AZ_{\tau\tau} Z_\chi\chi}{Z_\zeta\zeta} + Z_AZ_{\tau Q}}{\sqrt{-2\alpha + \frac{Z_\chi^2\chi^2}{Z_\alpha Z_\zeta\zeta}}}\mu^{\epsilon/2},
\end{eqnarray} \end{subequations}
in order to obtain a new expression for $\Gamma$ involving only terms linear in the sources. The renormalization factors ($Z$-factors) are calculable, see \cite{Dudal:2011gd} and underlying references like \cite{Verschelde:2001ia,Gracey:2002yt}. In the $\MSbar$ scheme and at one-loop, these $Z$-factors read in our current conventions as follows:
\begin{subequations} \label{Zs} \begin{gather}
	Z_A = 1+\frac{13}6 \frac{Ng^2}{16\pi^2} \frac2\epsilon \;, \qquad
	Z_g = 1-\frac{11}6 \frac{Ng^2}{16\pi^2} \frac2\epsilon \;, \qquad
	Z_{\varphi} = Z_g^{-1}Z_A^{-1/2} = 1+\frac34 \frac{Ng^2}{16\pi^2} \frac2\epsilon \;, \\
	Z_\zeta = 1-\frac{13}6 \frac{Ng^2}{16\pi^2} \frac2\epsilon \;, \qquad
	Z_\alpha = 1+\frac{35}{12}\frac{Ng^2}{16\pi^2} \frac2\epsilon \;, \qquad
	Z_\chi = 1 \;, \qquad
	Z_{\gamma^2} = Z_g^{-1/2}Z_A^{-1/4} = 1+\frac38 \frac{Ng^2}{16\pi^2} \frac2\epsilon \;, \\
	Z_{\tau\tau}=1-\frac{35}{12}\frac{Ng^2}{16\pi^2} \frac2\epsilon \;, \qquad
	Z_{\tau Q} = 0 \;, \qquad
	Z_{QQ} = Z_\varphi^{-1} = 1-\frac{3}{2}\frac{Ng^2}{16\pi^2\epsilon} \;.
\end{gather} \end{subequations}
Therefore, \eqref{W} can be rewritten as follows:
\begin{multline} \label{Final W}
	e^{-\Gamma(Q,\tau)} = \int [\mathcal D\Phi] [\mathcal D\sigma_1\mathcal D\sigma_2'] \exp\left[ -S_\text{GZ} - \int d^dx \left( \frac{\sigma_1^2}{2Z_\zeta}\left(1-\frac{\bar b^2}{\bar e^2} \frac{Z_\alpha}{Z_\zeta}\right) - \frac{\sigma_2'^2}{2Z_\alpha} - \frac{\bar b}{\bar e} \frac{\sigma_1\sigma_2'}{Z_\zeta} \right.\right. \\
	+ \left(\frac1{2Z_\zeta} \left(\bar a - \frac{\bar f\bar b}{\bar e}\right) \sigma_1 - \frac{\bar f}{2Z_\alpha} \sigma_2'\right) A^2 - \left(\frac{\bar b\bar d}{\bar e} \frac1{Z_\zeta} \sigma_1 + \frac{\bar d}{Z_\alpha} \sigma_2'\right) \overline\varphi\varphi \\
	\left.\left. + \frac{\bar a^2}{8Z_\zeta} (A^2)^2 - \frac1{2Z_\alpha} \left(\frac{\bar f}2 A^2 + \bar d \overline\varphi\varphi\right)^2 + \frac{\bar c}{Z_\zeta} \sigma_1\tau - \frac{\bar e}{Z_\alpha} \sigma_2' Q \right)\right]
\end{multline}
where $\sigma_2'$ is defined by
\begin{equation}\label{sigma3}
\sigma_2' = \sigma_2 - \frac{\bar b}{\bar e} \frac{Z_\alpha}{Z_\zeta} \sigma_1 .
\end{equation}

In this expression, all LCO parameters, sources and fields are now finite, and infinities are only present in the $Z$ renormalization factors, whether explicitly written or present in the coefficients $\bar a$ to $\bar f$. At one loop, $\chi=0$ and $Z_{\tau Q} = 0$ \cite{Dudal:2011gd}, which implies that $\bar{b}=\bar{f}=0$ and thus $\sigma_2' =\sigma_2$.%

In order to have an expression of the form $\frac{m^2}{2}A^2-M^2\bar{\varphi}\varphi$, we define the effective mass scales, $m^2$ and $M^2$, linked to $\Braket{AA}$ and $\Braket{\bar{\varphi}\varphi}$ respectively, by the classical (leading order in $g$) parts of the vacuum expectation values of the respective quadratic terms in the action \eqref{Final W}, that is:
\begin{subequations} \label{masses} \begin{gather}
	m^2 \equiv \left.\left(\frac1{Z_\zeta} \left(\bar a - \frac{\bar f\bar b}{\bar e}\right) \langle\sigma_1\rangle - \frac{\bar f}{Z_\alpha} \langle\sigma_2'\rangle\right)\right|_\text{leading} = \left. \frac1{\sqrt\zeta} \right|_\text{leading} \langle\sigma_1\rangle = \sqrt{\frac{13Ng^2}{9(N^2-1)}}\Braket{\sigma_1} \label{m^2 infinite} \\
	M^2 \equiv \left. \left(\frac{\bar b\bar d}{\bar e} \frac1{Z_\zeta} \langle\sigma_1\rangle + \frac{\bar d}{Z_\alpha} \langle\sigma_2'\rangle\right) \right|_\text{leading} = \left. \frac1{\sqrt{-2\alpha}} \right|_\text{leading} \langle\sigma_2'\rangle = \sqrt{\frac{35Ng^2}{48(N^2-1)^2}} \Braket{\sigma_2'} \label{M^2 infinite}
\end{gather} \end{subequations}
where the last equalities follow from $\alpha=\frac{\alpha_0}{g^2}=-\frac{24(N^2-1)^2}{35Ng^2}$ and $\zeta=\frac{\zeta_0}{g^2}=\frac{9(N^2-1)}{13Ng^2}$ \cite{Dudal:2011gd}.

Assuming the fields $\sigma_1$ and $\sigma_2'$ develop nonzero vacuum expectation values, we can compute these by means of Jackiw's background field method \cite{Jackiw:1974cv}. We replace these fields by a classical vacuum expectation value and a fluctuating quantum part, $\sigma \to \langle\sigma\rangle + \sigma$, ignore terms linear in the fields as these drop out when working around extrema, and we integrate out all the fluctuations. With this decomposition of the (auxiliary) fields, the quadratic part of the action (including only those $Z$-factors that are necessary for a one-loop computation) becomes
\begin{multline}
	\int d^dx \left( \frac{1}{2}A_\mu^a \left( - \delta_{\mu\nu} \partial^2 + \left(1-\frac1{\alpha_g}\right) \partial_\mu\partial_\nu \right) A_\nu^a + \bar c^a \partial^2 c^a + \overline\varphi_\mu^{ab} \partial^2 \varphi_\mu^{ab} - \overline\omega_\mu^{ab} \partial^2 \omega_\mu^{ab} \right. \\
	\left. - \gamma^2 g f^{abc} A_\mu^a (\varphi_\mu^{bc} + \overline\varphi_\mu^{bc}) - Z_{\gamma^2}^2 d ( N^2 -1) \gamma^4 + \frac{\sigma_1^2}{2Z_\zeta} - \frac{\sigma_2'^2}{2Z_\alpha} + \frac{m^2}2 A^2 - M^2 \overline\varphi\varphi \right) \;.
\end{multline}
Using the definitions \eqref{masses}, in addition to
\begin{subequations} \begin{gather}
	\bar{\varphi}_{\mu}^{ab}=U_{\mu}^{ab}+iV_{\mu}^{ab} \;, \qquad \varphi_{\mu}^{ab}=U_{\mu}^{ab}-iV_{\mu}^{ab} \;, \\
	P_{\mu\nu} \equiv(-\partial^2+M^2) \delta_{\mu\nu} \;, \\
	Q_{\mu\nu} \equiv \left[(-\partial^2+m^2)\delta_{\mu\nu}+\left(1-\frac1{\alpha_g}\right)\partial_{\mu}\partial_{\nu}\right] \;,
\end{gather} \end{subequations}
this quadratic part can be rewritten as
\begin{multline}
	\int d^dx \left( - Z_{\gamma^2}^2 d ( N^2-1) \gamma^4 + \frac{9(N^2-1)}{13Ng^2} \frac{m^4}{2Z_\zeta} - \frac{48(N^2-1)^2}{35Ng^2} \frac{M^4}{2Z_\alpha} \right. \\
	\left. + \frac12 A_\mu^a Q_{\mu\nu} A_\nu^a + \bar c^a \partial^2 c^a - U_\mu^{ab} P_{\mu\nu} U_\nu^{ab} - V_\mu^{ab} P_{\mu\nu} V_\nu^{ab} - \overline\omega_\mu^{ab} \partial^2 \omega_\mu^{ab} - 2 \gamma^2 g f^{abc} A_\mu^a U_\mu^{bc} \right) \;.
\end{multline}

Since the ghost fields $c$, $\bar{c}$, $\omega$, $\bar \omega$ appear uncoupled to other fields, they can be immediately integrated out, giving just an overall factor. The real bosonic fields $U$ and $V$ can be integrated out next, leading to:
\begin{equation}
\int[\mathcal{D}U,V]e^{-\int d^dx \left[-V_{\mu}^{ab}P_{\mu\nu}V_{\nu}^{ab}-U_{\mu}^{ab}P_{\mu\nu}U_{\nu}^{ab}-2g\gamma^2f^{abc}A_{\mu}^{a}U_{\mu}^{bc}\right]}
=\frac1{\det(P_{\mu\nu}\delta^{ac}\delta^{bd})}e^{-\int d^dx [Ng^2\gamma^4A_{\mu}^a P_{\mu\nu}^{-1} \delta^{ab}A_{\nu}^{b}]} \;,
\end{equation}
Introducing
\begin{equation}
	R_{\mu\nu} \equiv Q_{\mu\nu} + 2Ng^2\gamma^4 P_{\mu\nu}^{-1} = \left[\left(-\partial^2+m^2+\frac{2N\gamma^4g^2}{-\partial^2+M^2} \right)\delta_{\mu\nu} + \left(1-\frac1\alpha_{g}\right)\partial_{\mu}\partial_{\nu}\right] \;,
\end{equation}
we now also integrate over the gluon field $A_{\mu}$. The quadratic part of the action containing $A_{\mu}$ is
\begin{equation}
	 \int[\mathcal{D}A]  e^{-\frac{1}{2}\int d^dx A_{\mu}^aR_{\mu\nu}A_{\nu}^a} = \frac1{\sqrt{\det(R_{\mu\nu} \delta^{ab})}}
\end{equation}
As a result, the effective potential will be\footnote{We have tacitly removed the global volume factor everywhere.}
\begin{equation} \label{effpotbegin}
	\Gamma = - Z_{\gamma^2}^2 d ( N^2 -1) \gamma^4 + \frac{9(N^2-1)}{13Ng^2} \frac{m^4}{2Z_\zeta} - \frac{48(N^2-1)^2}{35Ng^2} \frac{M^4}{2Z_\alpha} + (N^2-1)^2 \operatorname{Tr}\ln P_{\mu\nu} + \frac{N^2-1}2{\rm Tr}\ln R_{\mu\nu} \;.
\end{equation}
The traces appearing in this expression are computed in the Appendix, see \eqref{finaleffpot}. Also defining $\lambda^4 \equiv 2Ng^2\gamma^4$, the one-loop renormalized effective potential of the Gribov--Zwanziger theory, refined with the condensates $\Braket{A_{\mu}^aA_{\mu}^a}$ and $\Braket{\bar{\varphi}_{\mu}^{ab}\varphi_{\mu}^{ab}}$, reads:
\begin{multline}\label{effective potential}
\Gamma(m^2,M^2,\lambda^4) = -\frac{2(N^2-1)}{Ng^2} \lambda^4\left(1 - \frac38 \frac{Ng^2}{16\pi^2}\right) + \frac{9(N^2-1)}{13Ng^2} \frac{m^4}2 - \frac{48(N^2-1)^2}{35Ng^2} \frac{M^4}2
	+ \frac{(N^2-1)^2}{8\pi^2} M^4 \left( - 1 + \ln\frac{M^2}{\bar\mu^2} \right) \\
	+ \frac{3(N^2-1)}{64\pi^2} \left(- \frac56 (m^4-2\lambda^4) + \frac{m^4+M^4-2\lambda^4}2 \ln\frac{m^2M^2+\lambda^4}{\bar\mu^4} \right. \\ \left. - (m^2+M^2) \sqrt{4\lambda^4-(m^2-M^2)^2} \arctan\frac{\sqrt{4\lambda^4-(m^2-M^2)^2}}{m^2+M^2} - M^4 \ln\frac{M^2}{\bar\mu^2} \right) \;.
\end{multline}

In \eqref{effective potential}, $m^2$ and $M^2$ are proportional to the vacuum expectation values $\Braket{\sigma_1}$ and $\Braket{\sigma_2'}$ of the auxiliary fields $\sigma_1$ and $\sigma_2'$ introduced through the Hubbard--Stratonovich transformations \eqref{HS transfo}, which may appear unphysical. However, acting with $\left. \frac{\delta}{\delta \tau} \right|_{\tau=Q=0}$ and $\left. \frac{\delta}{\delta Q} \right|_{\tau=Q=0}$  on \eqref{W} and \eqref{Final W} respectively, we get:
\begin{subequations} \label{link between condensates} \begin{gather}
\frac{1}{2}Z_{\tau}Z_A \Braket{A_{\mu}^a A_{\mu}^a} = \sqrt{\zeta} \mu^{-\epsilon/2} \Braket{\sigma_1} \;, \\
Z_Q Z_{\varphi} \Braket{\bar{\varphi}_\mu^{ac} \varphi_\mu^{ac}} = - \sqrt{-2\alpha} \mu^{-\epsilon/2} \Braket{\sigma_2'} \;.
\end{gather} \end{subequations}
The condensates $\Braket{\sigma_1}$ and $\Braket{\sigma_2'}$, and so the mass scales $m^2$ and $M^2$ entering in $\Gamma$, are thus directly related to the more intuitive BRST invariant condensates $\Braket{A_{\mu}^a A_{\mu}^a}_{\textrm{Landau}}\equiv \Braket{ A_\mu^h A_\mu^h}$ and  $\Braket{\bar{\varphi}_\mu^{ac} \varphi_\mu^{ac}}$ we were originally interested in, of which the LHS of \eqref{link between condensates} are the properly renormalized versions.

As expected, the condensation of the LCOs $A_{\mu}^a A_{\mu}^a$ and  $\bar{\varphi}_\mu^{ac} \varphi_\mu^{ac}$  modifies the energy density $\Gamma$ of the theory. The three first terms in the first line form the classical part of the potential while the rest of $\Gamma$, proportional to $g^2$ when we consider the $g$-dependence of $m^2$, $M^2$ and $\lambda^4$, is the one-loop quantum correction.

\section{Gap equation and minimization} \label{solve}
We now proceed to find the physical state of the vacuum. We need to solve the gap equation \eqref{gapG} while simultaneously minimizing with respect to $m^2$ and $M^2$.

The minus sign in front of $M^4$ in the second classical term\footnote{This is related to the sign of $\alpha_0$, which is ultimately dictated by the sign choice in the unity \eqref{poshs} we used.} obviously makes the classical potential unbounded from below and thus, unphysical. Our hope at this point was that the first order quantum correction could ``turn'' the potential, making it bounded from below --- and possessing one or several minima --- and thus physically meaningful at the quantum level. If it is the case, this would mean that this effective potential \eqref{effective potential} would have the remarkable property of being a pure quantum object, having no physical classical limit when $\hbar\rightarrow 0$.

A very qualitative asymptotic study gives $\Gamma\sim M^4\ln M^2$ for $M^2\rightarrow+\infty$ --- that is, the one-loop correction overtakes the classical term $-M^4$, as we hoped. Notice that this is qualitative at best, since taking field expectation values to infinity entails the presence of divergent logarithmic terms, making the efficacy of the perturbative computation of the  effective action again questionable. This issue is always present and has \emph{a~priori} nothing to do with the sign of the classical term. A full-fledged renormalization group improvement of the effective action goes far beyond the scope of the current paper, in particular since we are dealing with a multiscale problem. How to best deal with large expectation values in such cases is yet unsettled, see e.g.~\cite{Einhorn:1983fc,Bando:1992wy,Ford:1992mv,Chataignier:2018aud} for possible strategies, both old and new ones.

\subsection{Strategy to search for solutions}
To find the vacuum state of the theory, we need to solve the following gap equations:
\begin{equation}\label{gapEq}
\frac{\p \Gamma}{\p M^2}=0 \;, \qquad \frac{\p \Gamma}{\p m^2}=0  \;, \qquad \frac{\p \Gamma}{\p \lambda^4}=0 \;.
\end{equation}
As it is not possible to solve this very nonlinear system of equations by hand, we need to work numerically. In this case, it is necessary to make a choice for  the renormalization scale  $\bar\mu$ and  the coupling  $g$ before it is possible to start hunting for solutions. These choices are subject to several conditions:  as  we are working in a semiclassical approximation, we should  choose  $g$ to be sufficiently small that we can trust the perturbative approximation. The renormalization group then requires that $\bar\mu$ be sufficiently large, for we have  (at one loop in the $\overline{MS}$ scheme)
\begin{equation} \label{rge}
	\frac{Ng^2}{16\pi^2} = \frac1{\frac{11}3 \ln\frac{\bar\mu^2}{\lms^2}} \;.
\end{equation}
Furthermore, the scale $\bar\mu^2$ should be somehow ``close'' to the scales that appear in the logarithms (combinations of $m^2$, $M^2$, and $\lambda^2$), lest the logarithms appearing in higher-order corrections be too big to warrant a first-order approximation. In addition, the solution should be stable under variation of $m^2$ and $M^2$, as these will take the value that minimizes the action.\footnote{The value of $\lambda^2$ only needs to extremize the action, and indeed will normally maximize it. Although the latter might sound counterintuitive, it is actually a good sign. Indeed, we recall here that the original parameter $\gamma^2$ is the critical point coming from a saddle point evaluation \cite{Gribov:1977wm,Capri:2012wx}, so it better be corresponding to a maximum. The other parameters $m^2$ and $M^2$, however, need to be such that they minimize the action, for fixed $\gamma^2$. This means we need to verify, at the end, with the Hessian determinant criterion, that we have effectively found a minimum solution of the last 2 gap equations \eqref{gapEq}. } Finally, the existence of a nonzero solution for $\lambda$ is also a requirement, as otherwise the horizon condition would not be imposed,  and the formalism would be again plagued by Gribov copies .

To investigate this last requirement, let us write down the gap equation for the Gribov parameter $\lambda$, and use the renormalization group equation \eqref{rge} to eliminate the coupling $g$ in favor of $\bar\mu$ and $\lms$:
\begin{equation} \label{lambdagap}
	x \operatorname{arccot} x = \frac56 - \frac12 \ln\frac t{\bar\mu^4} + \frac{44}9 \ln\frac{\lms^2}{\bar\mu^2} \;,
\end{equation}
where we used the shorthands
\begin{equation}
	x = \frac{m^2+M^2}{\sqrt{4\lambda^4-(m^2-M^2)^2}} \;, \qquad t = m^2M^2 + \lambda^4 \;.
\end{equation}
In this equation \eqref{lambdagap}, there is still one choice we have to make: the value of $\bar\mu$. To simplify the computation, we will follow a backward approach: we will choose a value\footnote{This value can be any positive real number. If we choose a value larger than one, $x$ will be purely imaginary and $4\lambda^4<(m^2-M^2)^2$.} for $x\operatorname{arccot}x$, which determines $x$ and thus $\lambda$ as a function of the as yet undetermined $m^2$ and $M^2$. Next, we solve (still by hand) the gap equation \eqref{lambdagap} for $\bar\mu$ as a function of $m^2$ and $M^2$. Putting these solutions into the gap equations for $m^2$ and $M^2$, we can solve numerically for these two mass parameters. Plugging the solution back into the expressions we found for $\lambda$ and $\bar\mu$, we can determine the numerical values for these parameters as well.

Once a numerical solution has been found, we have to inspect its characteristics to see whether the solution is acceptable. In the $\MSbar$ scheme for $N=3$, it turns out that the effective coupling $Ng^2/16\pi^2$ is quite large for any value of $x\operatorname{arccot}x$ we may choose in \eqref{lambdagap}. The lowest value we obtained was $Ng^2/16\pi^2 = 1.7$. Other choices yielded either higher values of the coupling constant, or nonsensical negative $g^2$ values, or a saddlepoint when varying $m^2$ and $M^2$.

As the  difficulty  to find satisfactory solutions may be due to $\MSbar$ not being the  most convenient  subtraction scheme, we investigated other schemes. A scheme which  is often used is the momentum subtraction (MOM) scheme, as it can also be easily implemented on a lattice.  The relationship between this scheme and $\MSbar$ is computed in detail in \cite{Gracey:2011vw}. In our case, it turns out the first term in \eqref{effective potential} is to be replaced by
\begin{equation}
	-\frac{2(N^2-1)}{Ng^2} \lambda^4\left(1 - \left(\frac38 - \frac{5.233}N\right) \frac{Ng^2}{16\pi^2}\right) \;.
\end{equation}
Applying the procedure outlined above for $\MSbar$ still did not yield any satisfactory solutions, though.

\subsection{General subtraction scheme and lattice input}
In order to overcome these issues, we will ``optimize'' our one-loop effective action by considering it in a generic scheme. As is argued in \cite{Dudal:2005na},  we actually only need to parameterize two renormalization factors to change from the $\MSbar$ scheme to a general scheme since there are only two independent $Z$-factors in Landau gauge. In our case, it turns out to be most useful to consider $Z_g$ and $Z_{\gamma^2}$ as the independent $Z$-factors, and adapt the other $Z$-factors accordingly.   We also take into account that the LCO parameters always appear in combinations like $Z_\zeta \zeta$, the latter being renormalization group invariants themselves, see also the comments in \cite{Knecht:2001cc}. As a result, again only the first term in \eqref{effective potential} is modified, becoming
\begin{equation}
	-\frac{2(N^2-1)}{Ng^2} \lambda^4\left(1 - \left(\frac38 - b_0\right) \frac{Ng^2}{16\pi^2}\right) \;.
\end{equation}
where $b_0$ is a free parameter linked to the renormalization of the Gribov parameter $\gamma^2$, i.e.~to the finite part in the infinite renormalization factor $Z_{\gamma^2}$. The other freedom of scheme, lingering in the coupling constant renormalization, is yet invisible at one-loop order. As such, we can keep using the $\MSbar$ coupling.

With this general subtraction scheme, we again apply the steps outlined in the previous subsection to solve numerically for the effective mass scales $m^2$ and $M^2$ and the Gribov parameter $\gamma^2$, now  as functions  of the parameter $b_0$ in addition to the renormalization scale $\bar{\mu}$. Choosing the value of $b_0$ appropriately now does yield acceptable solutions. Now, however, we have too much freedom, and we need some extra criterion to fix $b_0$ again.

Applying the principle of minimal sensitivity \cite{Stevenson:1981vj} did not give anything useful: there was no optimal parameter choice. As such, we propose a different approach. The ultimate goal of this research program is to investigate what happens with the Gribov--Zwanziger theory at finite temperature, to investigate the response of the Green functions and their feedback on the deconfinement transition, if any, which can be investigated by including an appropriate temporal background \cite{Marhauser:2008fz,Braun:2007bx,Reinhardt:2013iia,Reinosa:2014ooa}, which allows to access the vacuum expectation value of the Polyakov loop. An important first step in this direction is to pinpoint a desirable $T=0$ vacuum state to start from. As such, we will benefit from lattice studies, of both SU(2) and SU(3) gauge theories, that have investigated how well a propagator of  the  Gribov type can describe the lattice gluon propagator, see \cite{Cucchieri:2011ig,Dudal:2018cli}. We will, however, not directly match our mass scales to the corresponding ones on the lattice, as this is a renormalization scheme and scale dependent operation. Instead we should use renormalization group invariant mass scales, which will be scale and scheme independent.

In the (R)GZ setting, there are two natural candidates, namely the set of complex conjugate poles of the gluon propagator \eqref{gluon_prop_faynman}. Next to being scale and scheme independent as pole masses\footnote{The generalization of the standard lore that a pole mass has  these  properties has been extended to the (R)GZ theory as well, see \cite{Capri:2016gut}.}, these quantities are even gauge parameter independent, thanks to the underlying BRST invariance, encoded in Nielsen identities \cite{Nielsen:1975fs,Piguet:1984js}. Practically speaking, we determine the complex conjugate poles of our propagator \eqref{gluon_prop_faynman} using the input of the one-loop effective potential, which depends on the 2 parameters $b_0$ and $\omu$, and we determine the latter two values by matching our estimate of these gluon poles with those as estimated from the lattice data, \cite{Dudal:2018cli} for $N=3$ and \cite{Cucchieri:2011ig} for $N=2$.

Let us first discuss the $N=3$ case. From the data given at the bottom of page 358  of ref. \cite{Dudal:2018cli} , righthand numbers, we can read off the denominator of the gluon propagator as $p^4 + 0.522 \;\mathrm{GeV}^2 p^2 + 0.2845\;\mathrm{GeV}^4$, from which the poles of the gluon propagator (which are our $x_\pm$, see \eqref{xpm}) are
\begin{equation}
	- p^2\bigg|_{\mathrm{pole}} = (0.26 \pm i 0.47) \mathrm{GeV}^2 = (5.2 \pm i 9.3) \lms^2 \;,
\end{equation}
where we used that $\lms = 0.224\;\mathrm{GeV}$ in $N=3$ pure Yang--Mills \cite{Boucaud:2008gn,Dudal:2017kxb}. A careful numerical analysis, following the above methodology, yields that for
\begin{equation}\label{opti}
	x\operatorname{arccot}x = 0.82 \;, \qquad b_0 = -3.42
\end{equation}
the equations allow for a solution with the gluon propagator pole at the right spot. In this solution we have
\begin{equation} \begin{gathered}
	\frac{g^2N}{16\pi^2} = 0.40 \;, \qquad \bar\mu = 1.41 \;\lms = 0.31\;\mathrm{GeV} \;, \\
	\Gamma = -24 \;\lms^4 = -0.059\;\mathrm{GeV}^4 \;, \qquad \lambda^4 = 28\;\lms^4 = 0.071\;\mathrm{GeV}^4 \;, \\
	m^2 = 2.6 \;\lms^2 = 0.13\;\mathrm{GeV}^2 \;, \qquad M^2 = 7.8 \;\lms^2 = 0.39\;\mathrm{GeV}^2 \;.
\end{gathered} \end{equation}
It turns out that the effective coupling constant is sufficiently small to attribute a qualitatively trustworthy meaning to our results. Furthermore we checked that the solution is a minimum under variation of $m^2$ and $M^2$ by computing the Hessian matrix at the minimum.

\begin{figure}
  \centering
  \includegraphics[width=0.3\textwidth]{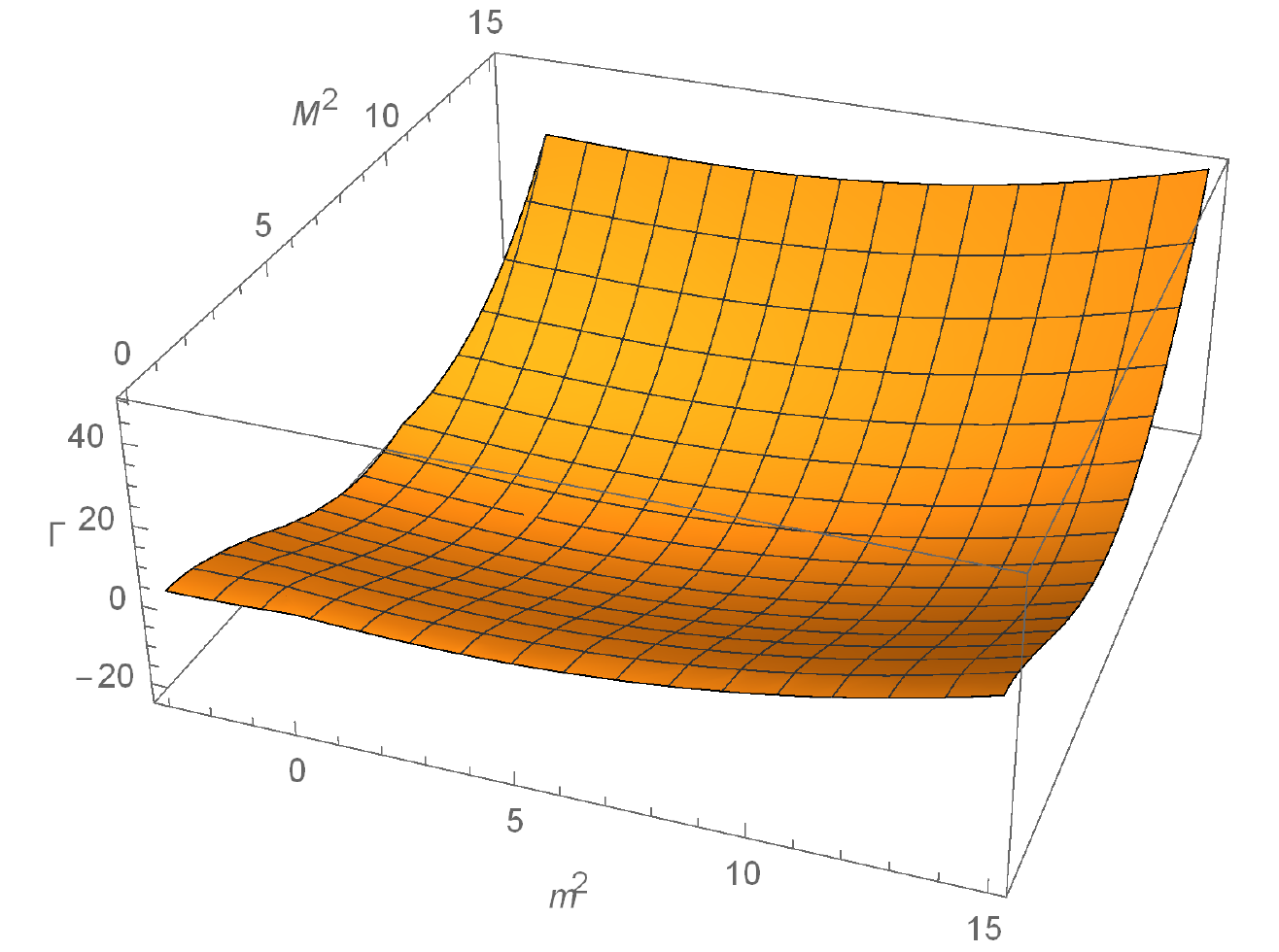}\qquad   \includegraphics[width=0.3\textwidth]{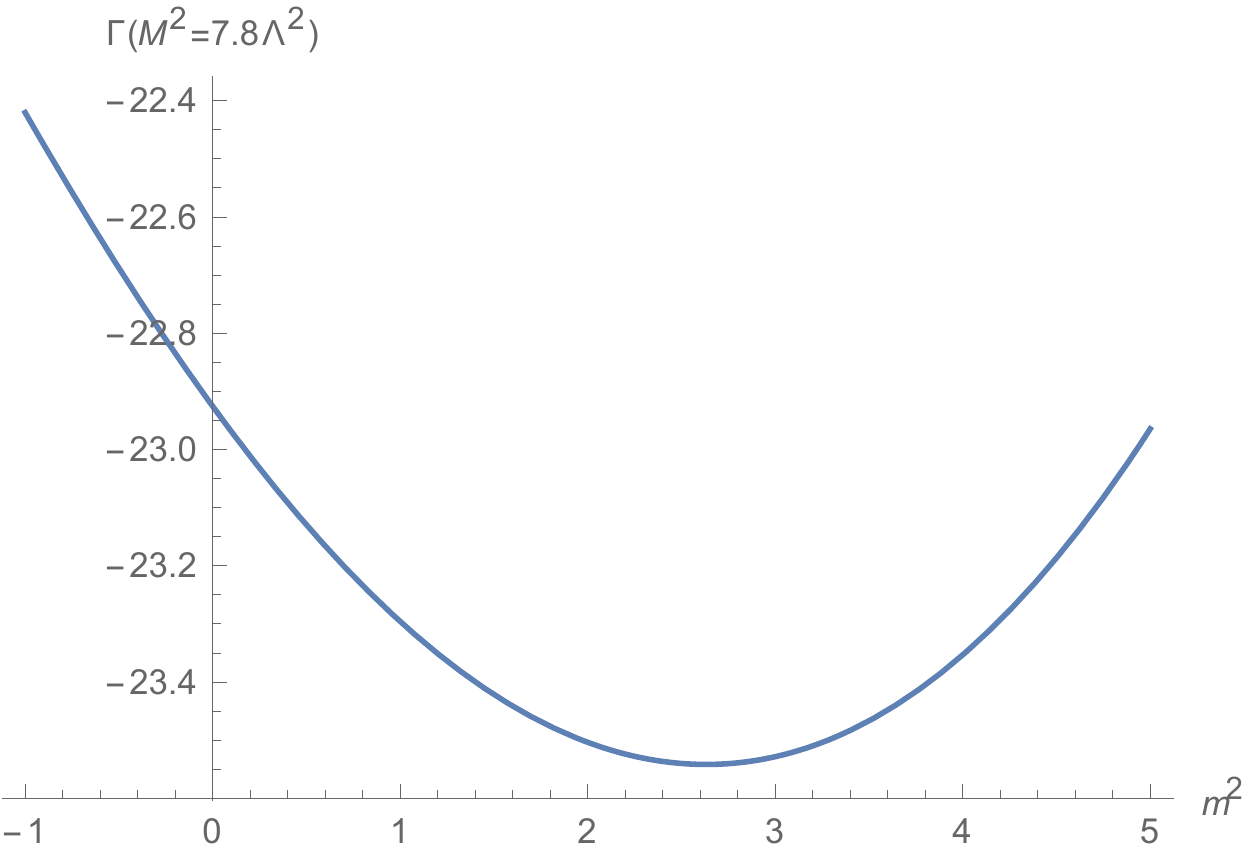}   \includegraphics[width=0.3\textwidth]{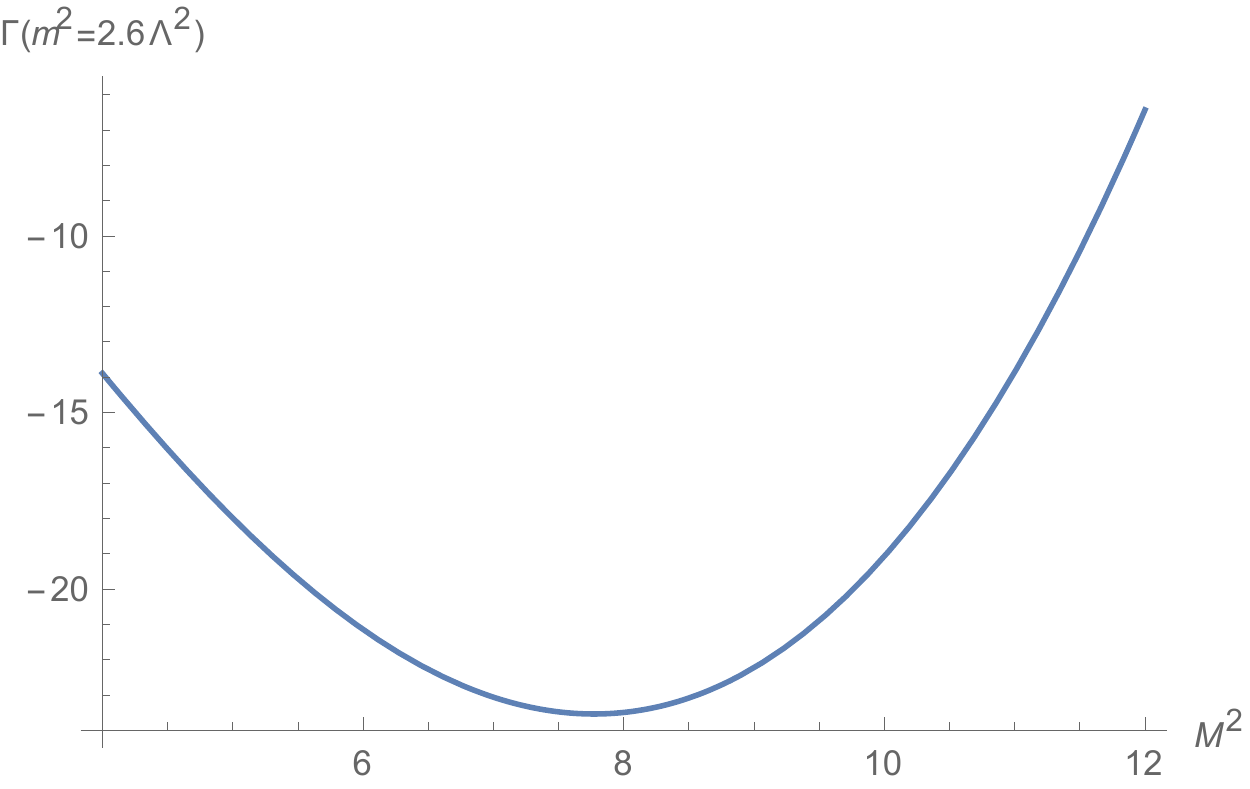}\\
  \caption{The effective action (left) and 2 slices thereof illustrating the minimum (right) ($N=3$ case, in units of $\lms$).}\label{fig1}
\end{figure}

 The main features of the above solution are captured in FIG.~\ref{fig1}.
It is instructive to notice that the vacuum energy is strictly negative, which shows that the non-perturbative vacuum in presence of the non-vanishing BRST invariant $d=2$ condensates is, at least up to one loop order, more stable than the ``pure'' GZ vacuum  $(m^2=M^2=0)$ , in which case it was already shown in \cite{Dudal:2005na} that the vacuum energy is always strictly positive, independent of the choice of massless renormalization scheme.
 In fact, at $M^2=m^2=0$, we have for any choice of scale or scheme that, at one-loop, $\frac{\p \Gamma}{\p m^2}<0$, $\frac{\p \Gamma}{\p M^2}<0$, so the pure GZ vacuum is indeed not stable.

\begin{figure}
  \centering
  \includegraphics[width=0.3\textwidth]{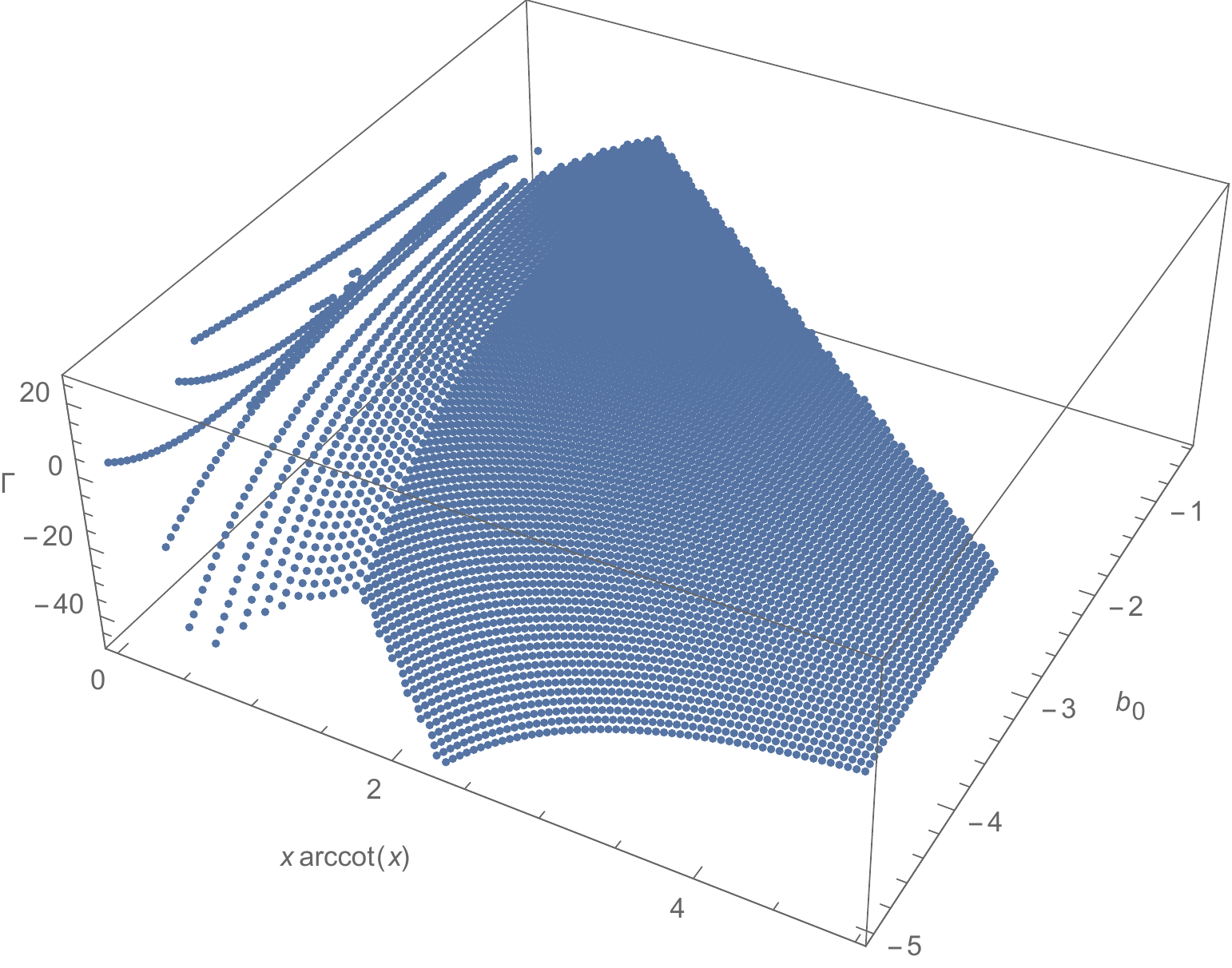}\\
  \caption{Dependence of $\Gamma$ on the renormalization scale and scheme ($N=3$ case, in units of $\lms$).}\label{fig2}
\end{figure}

 To get an idea of the sensitivity to the choice of scale and scheme, let us also present the results where the optimum values \eqref{opti} were increased with 25\%,
\begin{equation} \begin{gathered}
	\frac{g^2N}{16\pi^2} = 0.30 \;, \qquad \bar\mu = 1.57 \;\lms = 0.35\;\mathrm{GeV} \;, \\
	\Gamma = -65 \;\lms^4 = -0.16\;\mathrm{GeV}^4 \;, \qquad \lambda^4 = 27\;\lms^4 = 0.069\;\mathrm{GeV}^4 \;, \\
	m^2 = 2.3 \;\lms^2 = 0.11\;\mathrm{GeV}^2 \;, \qquad M^2 = 12.74 \;\lms^2 = 0.64\;\mathrm{GeV}^2 \;,
\end{gathered} \end{equation}
or decreased with 25\%,
\begin{equation} \begin{gathered}
	\frac{g^2N}{16\pi^2} = 0.66 \;, \qquad \bar\mu = 1.23 \;\lms = 0.27\;\mathrm{GeV} \;, \\
	\Gamma = -6 \;\lms^4 = -0.016\;\mathrm{GeV}^4 \;, \qquad \lambda^4 = 45\;\lms^4 = 0.11\;\mathrm{GeV}^4 \;, \\
	m^2 = 3.59 \;\lms^2 = 0.18\;\mathrm{GeV}^2 \;, \qquad M^2 = 4.35 \;\lms^2 = 0.22\;\mathrm{GeV}^2 \;.
\end{gathered} \end{equation}
For completeness, in FIG.~\ref{fig2}, we display the dependence of $\Gamma$ in the region with solutions, this to appreciate the fact that there is no optimal solution in the sense of minimal sensitivity.
This relatively strong dependence on the renormalization scale and scheme suggests that more stable results will, possibly, only be achieved within a full-blown higher loop study. This goes beyond the scope of the current paper. Two-loop computations in the (R)GZ context have not been established so far due to the large number of diagrams, not only caused by the extra vertices but also by, in particular, the several mixed propagators involving the gluon and extra GZ fields.

For SU(2),  we can be more brief. We use the results of \cite{Cucchieri:2011ig}. In Table IV of this paper, the poles of the propagator are given as
\begin{equation}
	- p^2\bigg|_{\mathrm{pole}} = (0.29 \pm i 0.66) \mathrm{GeV}^2 = (2.6 \pm i 6.0) \lms^2 \;,
\end{equation}
where we used that $\lms = 0.331\;\mathrm{GeV}$ in $N=2$ pure Yang--Mills \cite{Lucini:2008vi,Dudal:2017kxb}. We found that for
\begin{equation}
	x\operatorname{arccot}x = 0.74 \;, \qquad b_0 = -1.6
\end{equation}
the equations yielded a solution with the gluon propagator pole at the right spot. In this solution we have
\begin{equation} \begin{gathered}
	\frac{g^2N}{16\pi^2} = 1.24 \;, \qquad \bar\mu = 1.12 \;\lms = 0.37\;\mathrm{GeV} \;, \\
	\Gamma = -0.38 \;\lms^4 = -0.0046\;\mathrm{GeV}^4 \;, \qquad \lambda^4 = 9.1\;\lms^4 = 0.109\;\mathrm{GeV}^4 \;, \\
	m^2 = 2.3 \;\lms^2 = 0.25\;\mathrm{GeV}^2 \;, \qquad M^2 = 2.9 \;\lms^2 = 0.32\;\mathrm{GeV}^2 \;.
\end{gathered} \end{equation}
It turns out that the effective coupling constant is again a bit too high to really trust the  SU(2)  results;  we notice that the SU(2) and SU(3) results are in the same ballpark, related to the fact of course the input pole masses were rather similar.

\section*{Conclusion}
In this paper, we considered the recently introduced Gribov--Zwanziger action that implements the restriction of the gauge degrees of path integration to a smaller subregion in a way consistent with the linear covariant gauge condition while also removing a large set of Gribov (gauge) copies. We have explicitly constructed the one-loop effective potential for two $d=2$ condensates, related to BRST invariant operators. The latter property allows to carry out their computation (as well as that of the effective potential) in a specific gauge. We opted for Landau gauge, in which case the computation simplifies most. As the considered operators are local but composite, care is needed in how to construct the effective potential, related to renormalization (group) issues. We relied on the LCO (local composite operator) formalism of \cite{Verschelde:1995jj,Verschelde:2001ia,Dudal:2011gd}, which resolved all possible issues.

We computed the one-loop potential in a generic massless renormalization scheme, but were unable to pinpoint an optimal scheme, in the sense of minimal sensitivity. We therefore used lattice estimates for the set of complex conjugate poles of the gluon propagator, which are known to be renormalization group invariants. We then selected the (unique at the considered order) scheme in which the computed (tree level) complex conjugate poles match those lattice values. As such, we have identified a specific renormalization scheme to treat the divergences at zero temperature (the case considered here), upon which we can  build  in future work to discuss the interplay of condensates and Gribov gap equation with the temperature, with as ultimate goal to find out whether the GZ quantization can capture some essentials of the QCD thermodynamics and phase transitions, thereby putting on firmer footing preceding studies like \cite{Fukushima:2012qa,Fukushima:2013xsa,Canfora:2015yia,Canfora:2016ngn,Dudal:2017jfw,Kroff:2018ncl}.

The main result of this paper is the first explicit verification, albeit at one-loop order, that GZ dynamically transforms itself into RGZ thanks to the formation of nonperturbative $d=2$ mass scales, whilst respecting gauge and renormalization group invariance.  At the level of the propagators in a generic linear covariant gauge, our results are at least qualitatively consistent with lattice or other functional methods output.  This extends to vertices in the Landau gauge, for which many more results are available, see \cite{Mintz:2017qri} and references therein.

\section*{Acknowledgments}
D.~Vercauteren is grateful for the hospitality at KU Leuven, made possible through the Senior Fellowship SF/17/005.  C.P.~Felix is a PhD student
supported by the program Ci\^{e}ncias sem Fronteiras--CNPq, 234112/2014-0. L.~F.~Palhares thanks the hospitality at KU Leuven--Kulak where this work was initiated; she is partially supported by the Brazilian agencies CNPq (grants 454564/2014-7 and 305732/2016-1) and FAPERJ (grant E-26/203.197/2017) and is part of the project INCT-FNA (Process No. 464898/2014-5). This study was financed in part by the Coordena\c c\~ao de Aperfei\c coamento de Pessoal de N\'ivel Superior -- Brasil (CAPES) -- Financial Code 001 (M. N. F.). F.~Rondeau is supported by Ecole Normale Sup\'{e}rieure Paris-Saclay, and thanks the hospitality of KU Leuven campus Kulak Kortrijk.

\appendix
\section{Technical computational details}
\label{sect:append_a}
We compute the effective potential given in \eqref{effpotbegin}.

First of all there is the classical contribution
\begin{equation}
	- Z_{\gamma^2}^2 d (N^2-1) \gamma^4 + \frac{9(N^2-1)}{13Ng^2} \frac{m^4}{2Z_\zeta} - \frac{48(N^2-1)^2}{35Ng^2} \frac{M^4}{2Z_\alpha}
\end{equation}
Using the one-loop $Z$ factors given in \eqref{Zs}, this evaluates to
\begin{equation}
	-4(N^2-1)\gamma^4\left(1+\frac34 \frac{Ng^2}{16\pi^2} \frac2\epsilon -\frac38 \frac{Ng^2}{16\pi^2}\right) + \frac{9(N^2-1)}{13Ng^2} \frac{m^4}2 \left(1 + \frac{13}6 \frac{Ng^2}{16\pi^2} \frac2\epsilon\right) - \frac{48(N^2-1)^2}{35Ng^2} \frac{M^4}2 \left(1 - \frac{35}{12} \frac{Ng^2}{16\pi^2} \frac2\epsilon\right) \;.
\end{equation}

To compute the logarithmic traces of $P_{\mu\nu}$ and $R_{\mu\nu}$, we use the following well-known expression
\begin{equation}
	\operatorname{Tr} \ln(-\partial^2+\Delta) = - \frac1{(4\pi)^{d/2}} \Gamma(-\tfrac d2) \Delta^{d/2} = \frac{\Delta^2}{32\pi^2} \left( -\frac2\epsilon - \frac32 + \ln\frac\Delta{\bar\mu^2} \right) \;,
\end{equation}
where we used dimensional regularization ($d=4-\epsilon$) and the $\MSbar$ scheme. Using the fact that $\operatorname{tr}\delta_{\mu\nu} = d = 4-\epsilon$, we immediately find that
\begin{equation}
	\operatorname{Tr}\ln P_{\mu\nu} = \frac{M^4}{8\pi^2} \left( -\frac2\epsilon - 1 + \ln\frac{M^2}{\bar\mu^2} \right) \;.
\end{equation}
To compute the trace of the logarithm of $R_{\mu\nu}$, we first split the spectrum in one longitudinal polarization with eigenvalue $p^2/\alpha_g$, and $d-1$ transversal polarizations with eigenvalue
\begin{equation}
	p^2+m^2+\frac{2N\gamma^4g^2}{p^2+M^2} \;.
\end{equation}
As the longitudinal polarizations contribute nothing but an irrelevant constant, we only need to compute
\begin{equation} \begin{aligned}
	\operatorname{Tr}\ln R_{\mu\nu} =& (d-1) \operatorname{Tr} \ln \left( p^2+m^2+\frac{2N\gamma^4g^2}{p^2+M^2} \right) \\
	=& (d-1) \operatorname{Tr} \ln \bigg((p^2+m^2)(p^2+M^2)+2N\gamma^4g^2\bigg) - (d-1) \operatorname{Tr} \ln (p^2+M^2) \;.
\end{aligned} \end{equation}
Writing $2Ng^2\gamma^4 = \lambda^4$ and introducing the solutions of the equation $x^2+(M^2+m^2)x+M^2m^2+\lambda^4=0$, namely
\begin{equation} \label{xpm}
x_\pm = -\frac{1}{2}\left(m^2+M^2 \pm \sqrt{(m^2-M^2)^2-4\lambda^4}\right)
\end{equation}
this can be rewritten as
\begin{equation} \begin{aligned}
	\operatorname{Tr} \ln R_{\mu\nu} =& (d-1) \operatorname{Tr} \ln (p^2-x_+) + (d-1) \operatorname{Tr} \ln (p^2-x_-) - (d-1) \operatorname{Tr} \ln (p^2+M^2) \\
	=& \frac3{32\pi^2} (m^4-2\lambda^4) \left(-\frac2\epsilon-\frac56\right) + \frac3{32\pi^2} \left(x_+^2 \ln\frac{-x_+}{\bar\mu^2} + x_-^2 \ln\frac{-x_-}{\bar\mu^2} - M^4 \ln\frac{M^2}{\bar\mu^2} \right) \;.
\end{aligned} \end{equation}

Putting it all together, we find
\begin{multline} \label{finaleffpot}
	\Gamma = -4(N^2-1)\gamma^4\left(1 - \frac38 \frac{Ng^2}{16\pi^2}\right) + \frac{9(N^2-1)}{13Ng^2} \frac{m^4}2 - \frac{48(N^2-1)^2}{35Ng^2} \frac{M^4}2
	+ \frac{(N^2-1)^2}{8\pi^2} M^4 \left( - 1 + \ln\frac{M^2}{\bar\mu^2} \right) \\
	+ \frac{3(N^2-1)}{64\pi^2} \left(- \frac56 (m^4-2\lambda^4) + x_+^2 \ln\frac{-x_+}{\bar\mu^2} + x_-^2 \ln\frac{-x_-}{\bar\mu^2} - M^4 \ln\frac{M^2}{\bar\mu^2} \right) \;.
\end{multline}
The full effective potential is thus finite in the limit $\epsilon \rightarrow 0$ at first order in $g^2$, a nontrivial result and a strong indication that the computation is consistent. One can also verify the invariance of $\Gamma$ under the renormalization group.

\bibliographystyle{unsrt}
\bibliography{bibliography}

\end{document}